\documentclass{IET-Conf-Paper}

\begin{document}

\title{GreenBytes: Intelligent Energy Estimation for Edge-Cloud}
\author{Kasra Kassai\ad{1}, Tasos Dagiuklas, Satwat Bashir, Muddesar Iqbal}

\address{\add{}{School of Computer Science, London South Bank University, London, UK}
\email{\{kasra.kassai, tdagiuklas, bashis11, m.iqbal\}@lsbu.ac.uk}}


\begin{abstract}
This study investigates the application of advanced machine learning models, specifically Long Short-Term Memory (LSTM) networks and Gradient Booster models, for accurate energy consumption estimation within a Kubernetes cluster environment. It aims to enhance sustainable computing practices by providing precise predictions of energy usage across various computing nodes. Through meticulous analysis of model performance on both master and worker nodes, the research reveals the strengths and potential applications of these models in promoting energy efficiency. The LSTM model demonstrates remarkable predictive accuracy, particularly in capturing dynamic computing workloads over time, evidenced by low mean squared error (MSE) rates and the ability to closely track actual energy consumption trends. Conversely, the Gradient Booster model showcases robustness and adaptability across different computational environments, despite slightly higher MSE values. 
\end{abstract}

\maketitle

\section{Introduction}
\footnote{ This paper is a preprint of a paper submitted to the proceedings of 6G and Future Networks 2024  and is subject to Institution of Engineering and Technology Copyright. If accepted, the copy of record will be available at IET Digital Library}
The critical importance of energy estimation in computing stems from the pressing need for sustainable computing practices, a concept that aligns with the global drive towards environmental sustainability and reduced energy consumption. In the realm of computing, sustainable practices are pivotal not only for minimizing energy expenditure but also for reducing the carbon footprint associated with computing operations. Recent academic explorations, such as this study \cite{jagan2023}, highlight the growing field of sustainable computing. This research underscores the significance of optimizing computing practices to mitigate environmental impacts, particularly within power-intensive operations like large language model training. \cite{evolution}

The problem statement revolves around the current limitations in accurately estimating and thereby optimizing the energy consumption of computing resources \cite{estimation}. Traditional methods for energy estimation often fall short of capturing the dynamic nature of computing workloads and the complex interplay of hardware and software components. This inadequacy results in suboptimal energy usage and hinders the realization of truly sustainable computing environments. \cite{Energy-Aware, VM_placement, AdaptiveEdge}

Addressing these limitations, recent advancements propose more nuanced and adaptive approaches to energy estimation. For instance, the development of sophisticated models that incorporate real-time data and leverage artificial intelligence techniques to predict energy consumption more accurately. These models consider various factors such as workload characteristics, hardware efficiency, and operational settings to provide a more holistic and precise energy estimation\cite{Intel_Energy}.

The objective of our proof-of-concept experiment is to validate the efficacy of these advanced models in real-world scenarios. By doing so, we aim to demonstrate their potential to significantly improve energy efficiency and contribute to sustainable computing practices. Our approach seeks to not only address the existing gaps in energy estimation but also pave the way for more sustainable and cost-effective computing solutions. 

\section{Related Work}
The domain of energy estimation in computing, particularly focusing on LSTM networks and machine learning regression models, has witnessed significant advancements in recent years. These methods have been pivotal in enhancing the accuracy of energy consumption forecasts, which is crucial for developing sustainable and efficient computing systems\cite{Cloud-dc}. 

LSTM networks, a type of recurrent neural network (RNN), have been widely adopted for their capability to model time-series data effectively. For instance, \cite{lu2024novel} introduced a novel sequence-to-sequence-based deep learning model utilizing LSTM for multistep load forecasting, which significantly improves the prediction accuracy of energy demand in power system.

Similarly, the integration of machine learning regression models into energy estimation processes has shown promising results\cite{AI-assisted}. Rätz et al. \cite{RATZ2024} explored various machine learning techniques, including Gaussian Process Regression and LSTM networks, for optimizing energy consumption predictions. Their research highlights the versatility of machine learning models in handling complex datasets and producing reliable energy estimates. \cite{work-in-prog}

Despite these advancements, gaps remain in the literature, particularly in the scalability of these models  \cite{serverless, kepler, FCA, arm_energy_capture,KubeHICE} and their applicability to a broader range of computing environments. Many studies focus on specific applications, such as power load forecasting or battery state-of-charge estimation, with less emphasis on generalizing these models across different computing platforms and workloads. Additionally, there is a need for more comprehensive studies that combine LSTM networks with other machine learning techniques to enhance prediction accuracy further and reduce computational overhead. \cite{experimental-proof}

Our study aims to fill these gaps by developing an integrated approach that combines the predictive capabilities of LSTM networks with advanced regression models. By leveraging the strengths of both methodologies, we propose a more robust and versatile energy estimation model that can be adapted to various computing environments and workloads. This approach not only aims to enhance the accuracy of energy predictions but also contributes to the scalability and efficiency of sustainable computing practices.

Furthermore, our research addresses the challenge of real-time energy estimation, a relatively unexplored area in the existing literature. By developing a model that can provide accurate energy consumption estimates in real time, we seek to enable more dynamic and adaptive energy management strategies. This could significantly impact the development of sustainable computing systems, allowing for more precise control over energy usage and reducing the environmental footprint of computing operations.

\section{Methodology}

\subsubsection{Data Preprocessing}

The initial step in our methodology involves rigorous data preprocessing, essential for preparing the raw data for effective LSTM and Gradient Booster model training. Given the complexity and variability of computing environments, preprocessing plays a crucial role in ensuring the data's quality and relevance. The data is sourced from monitoring libraries such as mpstat \cite{mpstat}, which provide a comprehensive view of various CPU metrics that influence energy consumption.

\subsection{LSTM Model}
The preprocessed data, with a selected feature set, is then fed into the LSTM model for training. We adopt a training-to-validation data proportion of 0.2 to 0.8, ensuring a balanced approach to model learning while retaining a substantial dataset for validation purposes. This ratio allows for comprehensive model training on a diverse dataset, promoting the development of a robust model capable of generalizing well to unseen data.

The LSTM model is trained to predict energy consumption based on the selected features, with the architecture designed to capture long-term dependencies in the data. This capability is crucial for accurately forecasting energy usage patterns, which are often influenced by prolonged system behaviours and workload characteristics.

\subsection{Gradient Booster Model}

The gradient booster model employed in our study utilizes decision trees as base learners within an ensemble technique to enhance predictive accuracy and robustness for energy consumption estimation\cite{priyadarshini2022machine}. It excels in modelling complex, non-linear relationships between various CPU metrics and energy usage, making it well-suited for the intricate dynamics of computing environments. The model's iterative correction of errors and focus on influential features ensure a high level of generalization to new data, providing reliable estimates of energy consumption. Additionally, it offers valuable insights into feature importance, guiding further energy optimization efforts.

\subsubsection{Experiment Setup and data pipeline}

Normalization: Scaling the data to a standard range to facilitate model training by speeding up the convergence.
Cleaning: Removing outliers and handling missing values to ensure the integrity and reliability of the training data.
Feature Selection: Identifying the most relevant features that significantly impact energy consumption. In our study, selected features include CPU utilization at both the system and user levels, context switches per second, and other pertinent CPU metrics such as CPU frequency and idle states.
Feature Selection
The selection of features for the LSTM model is critical, focusing on those that directly influence energy consumption:

CPU Utilization: Both system and user-level utilization metrics are considered, as they directly correlate with the energy consumed by computing processes.

Context Switches per Second: This metric indicates the frequency of switching between processes, which can significantly impact energy usage due to the overhead associated with context switching.

Other CPU Metrics: Additional metrics obtained from Powertop, including CPU interrupt request signal (IRQ) and idle states, are incorporated into the feature set. These metrics provide insights into the efficiency and energy demands of the CPU under various workloads.

\subsubsection{Test Bed Setup}
Our experimental setup comprises three Intel-based Ubuntu box machines, chosen for their relevance to typical computing environments and the availability of detailed monitoring data through libraries like Powertop or mpstat. This data monitoring is only possible through intel RAPL\cite{Intel_RAPL} which has provided the target values or what we label as actual consumption in this study. These machines serve as the test bed for collecting the necessary data for model training and evaluation.

\subsubsection{Data Collection and Storage}
The data collection process leverages the Powertop library to gather detailed metrics related to CPU usage and other system behaviors influencing energy consumption. This data is meticulously recorded and stored in a dictionary format, facilitating efficient preprocessing and feature extraction. The structured format ensures that each metric is accurately associated with its corresponding time stamp and system state, laying a solid foundation for subsequent analysis.

\subsection{Results and Analysis:}

The objective of the study was to evaluate the effectiveness of two distinct machine learning models, LSTM and Gradient Booster, in predicting energy consumption within a Kubernetes 1.28 cluster environment. The analysis was performed across master and worker nodes, each representing unique computational workloads and energy usage patterns.

\subsubsection{LSTM Model Performance}

\paragraph{Master Node}

\begin{figure}
    \centering
    \includegraphics[width=1\linewidth]{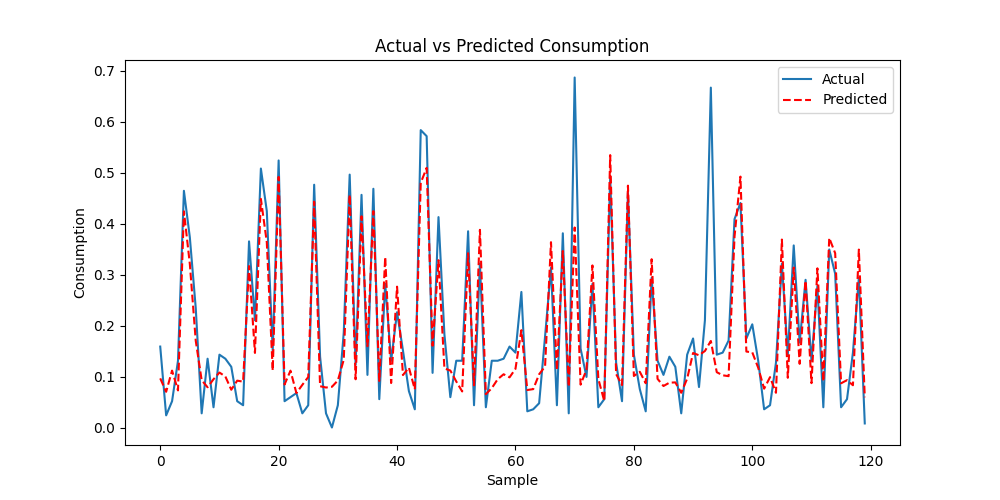}
    \caption{Actual vs Predicted Consumption on Master node running LSTM model}
    \label{fig:label1}
\end{figure}

On the master node, the LSTM model demonstrated promising results. The model’s loss over 100 epochs showed a rapid decline initially, suggesting a swift learning curve, and then plateaued, indicating that it had effectively captured the temporal patterns in the data without overfitting. This was reflected in the model's stable and low test loss throughout the training process. The subsequent analysis of actual vs. predicted energy consumption, measured in kilowatt-hours (kWh), revealed a strong correlation between the model's forecasts and the actual energy usage, albeit with slight deviations at peaks of consumption. These findings imply that the LSTM model is highly proficient in predicting energy consumption trends and can be instrumental in devising energy-efficient strategies shown in Figure \ref{fig:label1}.

\paragraph{Worker Nodes}

\begin{figure}
    \centering
    \includegraphics[width=1\linewidth]{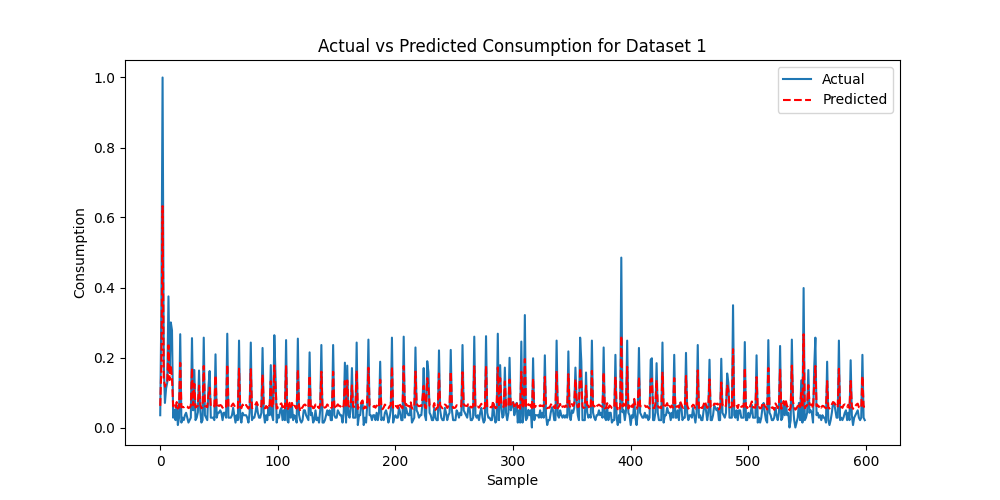}
    \caption{Actual vs Prediction Consumption for worker 1 using LSTM model trained on Master node}
    \label{fig:label2}
\end{figure}

\begin{figure}
    \centering
    \includegraphics[width=1\linewidth]{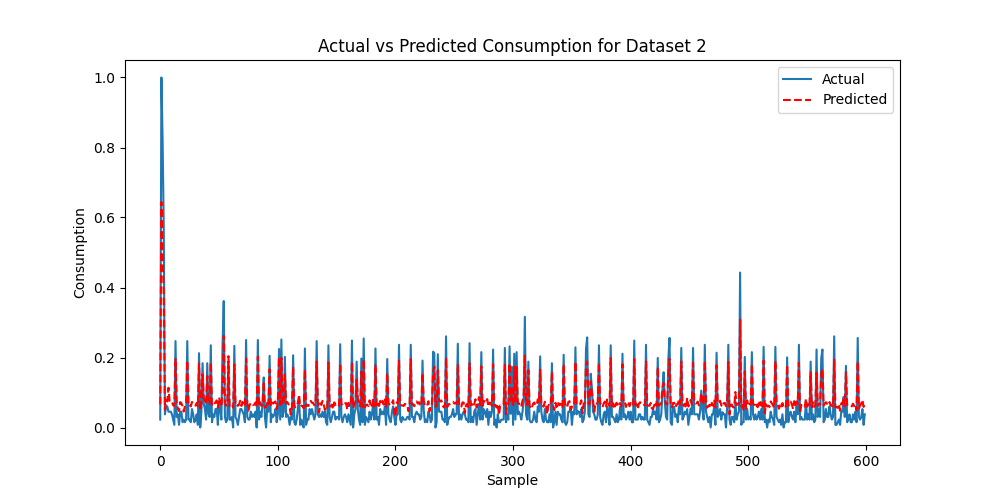}
    \caption{Actual vs Prediction Consumption for worker 2 using LSTM model trained on Master node}
    \label{fig:label3}
\end{figure}

On worker nodes, the LSTM model’s performance was equally impressive. For Worker Node 1, the MSE was recorded at 0.0018, denoting a high level of prediction accuracy. The model closely followed the actual consumption pattern over time. Worker Node 2 displayed a slightly higher MSE of 0.0019 but still showcased consistent accuracy. The ability of the LSTM model to closely mirror the energy consumption patterns on both nodes validates its suitability for real-time energy monitoring in distributed computing systems. It confirms the model's potential for scalability across various computational nodes, making it a valuable asset for Kubernetes orchestration and sustainable energy practices as shown in Figures \ref{fig:label2} and \ref{fig:label3}.


\subsubsection{Gradient Booster Model Performance}
\paragraph{Master Node}

\begin{figure}
    \centering
    \includegraphics[width=1\linewidth]{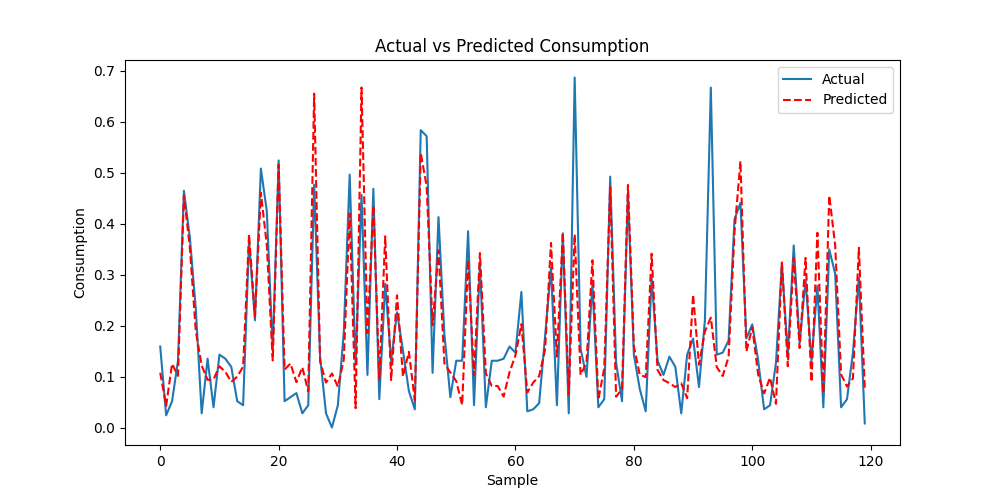}
    \caption{Actual vs Predicted Consumption on Master node running Gradient Booster model}
    \label{fig:label4}
\end{figure}

The Gradient Booster model on the master node showed an MSE that decreased sharply in the initial iterations and then stabilized, indicating an efficient learning of the training data and a strong generalization capability. The visual comparison between the actual and predicted energy consumption illustrated a high level of precision, with the model capturing complex patterns of energy usage in the master node's environment. These results underscore the model's potential in refining energy management strategies, providing a dependable tool for system administrators to enhance energy efficiency in computing clusters illustrated in \ref{fig:label4}.

\begin{figure}
    \centering
    \includegraphics[width=1\linewidth]{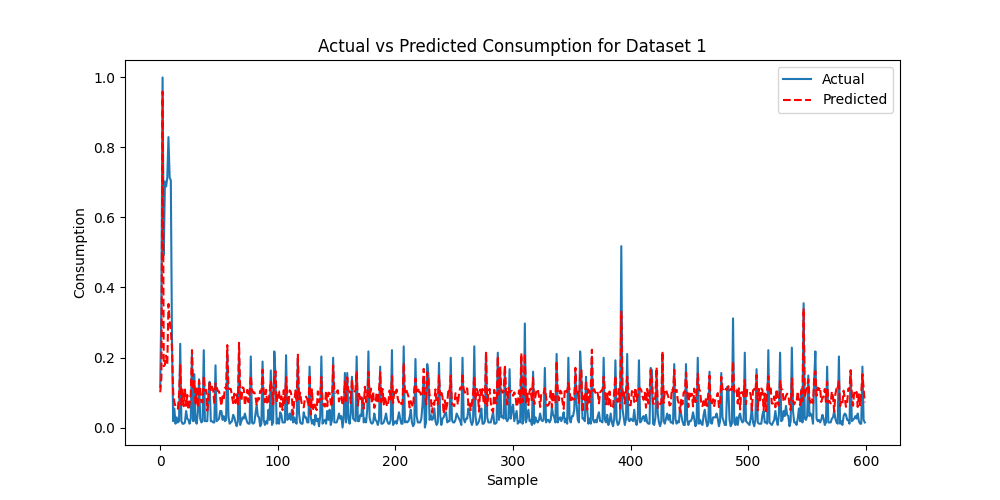}
    \caption{Actual vs Prediction Consumption for worker 1 using Gradient Booster model trained on Master node}
    \label{fig:label5}
\end{figure}

\begin{figure}
    \centering
    \includegraphics[width=1\linewidth]{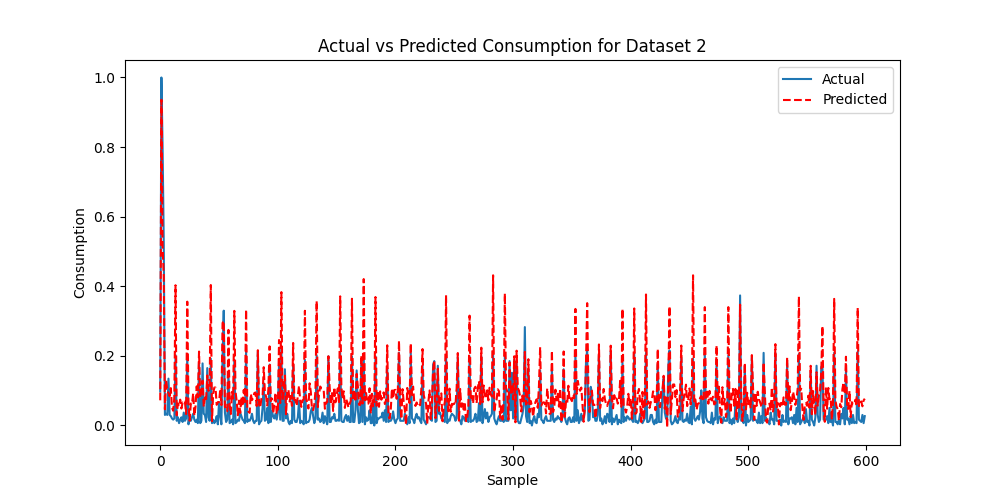}
    \caption{Actual vs Prediction Consumption for worker 2 using Gradient Booster model trained on Master node}
    \label{fig:label6}
\end{figure}

\paragraph{Worker Nodes}
For Worker Node 1, the Gradient Booster model recorded an MSE of 0.00677, reflecting a satisfactory predictive performance. Although this was higher compared to the LSTM model, it still captured the energy demand fluctuations. Worker Node 2 showed an improved MSE of 0.00450, indicating the model's adaptability to different node workloads and operational conditions. The more refined prediction accuracy suggests that the Gradient Booster model could be particularly suited to certain node usage patterns, potentially due to more consistent or predictable workloads illustrated in \ref{fig:label5} and \ref{fig:label6}.

\subsubsection{Comparative Analysis}
A comparative analysis of both models highlights that the LSTM model generally provided a closer fit to the actual data, as evidenced by the lower MSE values. However, the Gradient Booster model demonstrated a robust and adaptable performance across different worker nodes, suggesting that it could be better suited for environments where workloads and operational conditions vary significantly.

The performance of both models is indicative of their potential to inform and optimize energy management strategies within large-scale Kubernetes clusters \cite{kubernetes2019kubernetes}. The high degree of accuracy in the LSTM model’s predictions is particularly promising for tasks requiring substantial computational power, where energy demand directly correlates with workload intensity. The Gradient Booster model’s robust predictions across various nodes suggest its potential as a tool for energy estimation that could contribute significantly to the implementation of sustainable practices in distributed computing systems.

\section{Conclusion}
In this study, we investigated the viability of employing advanced machine learning models, specifically Long Short-Term Memory (LSTM) networks and Gradient Booster models, for the precise estimation of energy consumption within a Kubernetes cluster environment. Our analysis meticulously assessed the performance of these models across both master and worker nodes, unveiling their strengths and potential applications in enhancing the sustainability of computing practices.

The LSTM model demonstrated remarkable accuracy in forecasting energy consumption patterns, a testament to its ability to understand and predict the dynamic nature of computing workloads over time. This model's proficiency was particularly evident in its low mean squared error (MSE) rates and its capacity to closely track actual energy usage trends, thereby offering a solid foundation for the development of energy-efficient strategies. Such predictive capabilities are crucial for managing the energy demands of intensive computational tasks, enabling more environmentally conscious computing operations.

Conversely, the Gradient Booster model, while exhibiting slightly higher MSE values in comparison, still provided valuable insights into energy consumption patterns. Its performance underscores the adaptability of this model to varying computational environments and workloads, suggesting its utility in scenarios where operational conditions and workloads are less predictable. The Gradient Booster model's strength lies in its robustness and generalization ability, making it an essential tool for system administrators aiming to optimize energy efficiency across diverse computing platforms.

The comparative analysis of both models reveals that while the LSTM model offers a closer fit to actual energy consumption data, the Gradient Booster model's adaptability makes it equally invaluable. These findings highlight the complementary nature of these models in advancing sustainable computing practices. By integrating these models into energy management systems, stakeholders can leverage their predictive accuracy to implement more dynamic and responsive energy optimization strategies. This approach not only enhances the sustainability of computing environments but also contributes to the broader goal of reducing the environmental impact of technology operations.

The compiled results section, while providing a comprehensive view of the models' performance, suggests areas for further examination. A more detailed comparative analysis between the LSTM and Gradient Booster models could elucidate their relative strengths and limitations, offering deeper insights into their applicability in diverse computing contexts. Additionally, incorporating a broader range of performance metrics beyond MSE, such as R-squared or Mean Absolute Error (MAE), could furnish a more nuanced evaluation of the models. Discussing the implications of predictive accuracy on actual energy management strategies and potential cost savings would enrich the findings, as would addressing the models' scalability and adaptability to different computing environments. These considerations could reinforce the study's conclusions and provide a clearer pathway for future research.

In conclusion, our research contributes to the growing body of knowledge on sustainable computing by demonstrating the effectiveness of LSTM and Gradient Booster models in energy consumption estimation. As the demand for computing resources continues to rise, the insights gleaned from our study are pivotal for the development of more energy-efficient and environmentally friendly computing systems. Future work should explore the integration of these models into real-world computing environments, further refining their predictive capabilities and extending their application to a wider range of computing contexts. This endeavour will play a critical role in the ongoing pursuit of sustainability within the realm of computing, aligning technological advancement with environmental stewardship.

\section{References}

\bibliographystyle{ieeetr}
\bibliography{ref}
\end{document}